%% file: g_recoil_2p3.tex
\documentclass[preprint,aps,pra,showpacs,floatfix]{revtex4-1}
\usepackage[utf8]{inputenc}
\usepackage[OT1]{fontenc}
\usepackage{amsmath}
\usepackage{amsfonts}
\usepackage{amssymb}
\usepackage{bm}
\usepackage{graphicx}
\usepackage[left=2cm,right=2cm,top=2cm,bottom=2cm]{geometry}
\allowdisplaybreaks

\usepackage{longtable}
\usepackage{dcolumn}
\usepackage{siunitx}
\usepackage{multirow}

\newcommand{\veps}{\varepsilon}
\newcommand{\balpha}{\bm{\alpha}}
\newcommand{\bnabla}{\bm{\nabla}}

\newcommand{\bfp}{{\bf p}}
\newcommand{\bfA}{{\bf A}}
\newcommand{\bfD}{{\bf D}}
\newcommand{\bfr}{{\bf r}}

\newcommand{\be}{\begin{eqnarray}}
\newcommand{\ee}{\end{eqnarray}}

\newcommand{\psum}{\sideset{}{'}\sum}

\newcommand{\la}{\langle}
\newcommand{\ra}{\rangle}

\usepackage{color}
\definecolor{BLUE}{rgb}{0.0,0.0,1.0}

\definecolor{GREEN}{rgb}{0.0,0.7,0.0}

\usepackage{soul}
\soulregister\cite7
\soulregister\ref7

\begin{document}

\title{Relativistic calculation of the nuclear recoil effect on the $g$ factor of the $^2P_{3/2}$ state in highly charged B-like ions}

\author{A. V. Malyshev, D. A. Glazov, I. A. Aleksandrov, I. I. Tupitsyn, and V. M. Shabaev}
\affiliation {Department of Physics, St.~Petersburg State University, 
Universitetskaya 7/9, 199034 St.~Petersburg, Russia}


\begin{abstract}

The nuclear recoil effect on the $^2 P_{3/2}$-state $g$ factor of B-like ions is calculated to first order in the electron-to-nucleus mass ratio $m/M$ in the range $Z=18$--$92$. The calculations are performed by means of the $1/Z$ perturbation theory. Within the independent-electron approximation, the one- and two-electron recoil contributions are evaluated to all orders in the parameter $\alpha Z$ by employing a fully relativistic approach. The interelectronic-interaction correction of first order in $1/Z$ is treated within the Breit approximation. Higher orders in $1/Z$ are partially taken into account by incorporating the screening potential into the zeroth-order Hamiltonian. The most accurate to date theoretical predictions for the nuclear recoil contribution to the bound-electron $g$ factor are obtained.

\end{abstract}

\maketitle

\section{Introduction}

High-precision measurements of the $g$ factor of highly charged ions~\cite{Haffner:2000:5308,Verdu:2004:093002,Sturm:2011:023002,Sturm:2013:030501_R,Wagner:2013:033003,Sturm:2014:467,Kohler:2016:10246,Arapoglou:2019:253001,Glazov:2019:173001} have demonstrated unique opportunities for determination of the fundamental constants and verification of the bound-state quantum electrodynamics (QED)~\cite{Shabaev:2015:031205,Sturm:2017:4,Shabaev:2018:60}. While H-like ions are the simplest ones for theoretical computations, few-electron systems, e.g., Li- and B-like ions, turn out to be indispensable for the proposed scenarios of determining the fine-structure-constant $\alpha$~\cite{Shabaev:2006:253002,Volotka:2014:023002,Yerokhin:2016:100801}. Tests of bound-state QED in the strong-coupling regime, both within the Furry picture~\cite{Shabaev:2002:062104} and beyond~\cite{Malyshev:2017:765}, are also based on studies of the $g$ factor of few-electron ions. Motivated by these long-term goals, the high-precision experiments with Li-like ions~\cite{Wagner:2013:033003,Kohler:2016:10246,Glazov:2019:173001} were performed. The agreement established between theory and experiment on the level of $10^{-9}$ manifests the most stringent test of the many-electron QED effects in the presence of an external magnetic field~\cite{Volotka:2014:253004,Yerokhin:2017:062511,Glazov:2019:173001}. The ground-state $g$ factor of B-like argon was recently measured with an accuracy of $0.9\times 10^{-9}$~\cite{Arapoglou:2019:253001}. It yielded perfect agreement with the theoretical value, which has, however, a much larger uncertainty, $0.6\times 10^{-6}$~\cite{Glazov:2013:014014,Shchepetnov:2015:012001,Agababaev:2018:012003,Arapoglou:2019:253001}. The first experimental $g$-factor value for the $2P_{3/2}$ state which was accurate enough to probe the QED effects was obtained in a spectroscopic measurement of a fine-structure transition in B-like argon~\cite{SoriaOrts:2007:052501}. An analogous experiment conducted recently with a greater accuracy provided the value $g_\mathrm{exp}[{}^{40}_{18}\mathrm{Ar}^{13+}(^2P_{3/2})]=1.33214(15)$~\cite{Egl:2019:123001}, which is in agreement with theory~\cite{Glazov:2013:014014,Shchepetnov:2015:012001,Agababaev:2019:Xray_inpress}. Moreover, the anticipated experiments~\cite{Lindenfels:2013:023412,Vogel:2019:1800211,Sturm:2019:1425} may bring the precision of the $2P_{3/2}$-state $g$ factor to the level taking place for the ground state of H-, Li-, and B-like ions, i.e., $10^{-9}$ or better. Thus, further theoretical studies of the $g$ factor of both $2P_{1/2}$ and $2P_{3/2}$ states are highly motivated.

The first measurement of the $g$-factor isotope shift in Li-like calcium~\cite{Kohler:2016:10246} has stimulated recent investigations of the nuclear recoil correction to the $g$ factor of highly charged Li-like ions~\cite{Shabaev:2017:263001,Malyshev:2017:765,Shabaev:2018:032512}. A rigorous evaluation of the nuclear recoil effect within the QED approach developed in Ref.~\cite{Shabaev:2001:052104} was first performed for the $1s$ state in Ref.~\cite{Shabaev:2002:091801}. In Refs.~\cite{Kohler:2016:10246,Shabaev:2017:263001,Malyshev:2017:765,Shabaev:2018:032512} it was extended to the $2s$ state. In our recent study~\cite{Malyshev:2019:G_rec_H:preprint}, the recoil effect was considered for the $1s$, $2s$, $2p_{1/2}$, and $2p_{3/2}$ states of H-like ions in the low-$Z$ region ($Z=1$--$20$). At the same time, in Refs.~\cite{Shabaev:2017:263001,Shabaev:2018:032512} the interelectronic-interaction contribution to the nuclear recoil effect in Li-like ions was evaluated within the Breit approximation derived from the fully relativistic theory~\cite{Shabaev:2001:052104}.


Later on, these studies have been extended to the ground $^2P_{1/2}$ state of B-like ions. The Breit-approximation values up to the first order in $1/Z$ have been obtained in the wide $Z$ range ($Z=10$--$92$) in Refs.~\cite{Glazov:2018:457,Aleksandrov:2018:062521} and complemented by the second- and higher-order contributions in $1/Z$ in Ref.~\cite{Glazov:2019:G_rec_B:preprint_sumbit}. QED calculations to all orders in $\alpha Z$ were carried out in Refs.~\cite{Aleksandrov:2018:062521,Glazov:2019:G_rec_B:preprint_sumbit} for the $^2P_{1/2}$ state. As for the $^2P_{3/2}$ state, only the zeroth-order Breit-approximation results are available in the range $Z=10$--$20$~\cite{Glazov:2013:014014,Agababaev:2019:Xray_inpress}. The first-order $1/Z$ correction was calculated in the nonrelativistic approximation only for $Z=18$~\cite{Shchepetnov:2015:012001}.

In the present study, we fill the gap related to the $^2P_{3/2}$ state of B-like ions by calculating both the Breit and QED parts of the nuclear recoil correction to the $g$ factor. The Breit-approximation results include the zeroth and first orders in $1/Z$ obtained with a screening potential, so they partially take into account the higher-order contributions. The one- and two-electron QED terms are evaluated to all orders in $\alpha Z$ and to zeroth order in $1/Z$ with a screening potential. As a result, the most accurate to date theoretical values of the nuclear recoil correction to the $^2P_{3/2}$-state $g$ factor of B-like ions in the range $Z=18$--$92$ are presented including B-like argon ion, which is a subject of the current and future experimental investigations. The results obtained provide an important contribution to further improvements of the theoretical $g$-factor value.

Relativistic units ($\hbar=1, c=1$) and Heaviside charge unit ($e^2=4\pi\alpha$, $e<0$) are employed throughout the paper, $\mu_0 = |e|/2m$ is the Bohr magneton.

\section{Theoretical methods}

A quantum electrodynamic theory of the nuclear recoil effect on the atomic $g$ factor to first order in the electron-to-nucleus mass ratio $m/M$ was formulated in Ref.~\cite{Shabaev:2001:052104}. In the case of one electron over closed shells, the complete $\alpha Z$-dependent formula for the nuclear recoil correction to the bound-electron $g$ factor reads
\begin{equation}
\Delta g = \frac{1}{\mu_0 m_a}\frac{1}{M}\frac{i}{2\pi}
\int \limits_{-\infty}^{\infty} d\omega\;
\left[\frac{\partial}{\partial {\mathcal H}}
\la \tilde{a} | [p^k-D^k(\omega)+eA_{\rm cl}^k]
                \tilde{G}(\omega+\tilde{\veps}_a)
                [p^k-D^k(\omega)+eA_{\rm cl}^k] \tilde{a} \ra
\right]_{{\mathcal H}=0}\, .
\label{eq:g_general}
\end{equation}
The wave function $|\tilde{a}\rangle$ of the valence electron is an eigenstate of the Dirac Hamiltonian for a spherically symmetric binding potential $V(r)$ in the presence of the external homogeneous magnetic field $\boldsymbol{\mathcal{H}}=\mathcal{H}{\bf e}_z$,
\begin{align}
\label{eq:Dirac_magn}
\tilde{h}_{\rm D} &= h_{\rm D} + \mu_0\mathcal{H}{m}\,[ \bfr \times \balpha ]_z \, , \\
\label{eq:Dirac}
h_{\rm D} &= -i \balpha \cdot \bnabla + \beta m + V(r) \, ,  \\
\label{eq:eigen_magn}
\tilde{h}_{\rm D} | \tilde{a} \rangle &= \tilde{\veps}_a | \tilde{a} \rangle \, .
\end{align}
In Eq.~(\ref{eq:g_general}) $m_a$ is the angular momentum projection, $\bfp=-i\bnabla$ is the momentum operator, $\bfA_{\rm cl}(\bfr) = [\boldsymbol{\mathcal{H}} \times \bfr ] /2$ is the classical vector potential, and the vector $D^k(\omega)=-4\pi\alpha Z\alpha^l D^{lk}(\omega)$ arises from the transverse part of the photon propagator in the Coulomb gauge,
\begin{equation}\label{eq:D_lk}
D^{lk}(\omega,{\bf r}) = 
-\frac{1}{4\pi}\left[\frac
{\exp{(i|\omega|r)}}{r}\delta_{lk} +\nabla^{l}\nabla^{k}
\frac{(\exp{(i|\omega|r)}
-1)}{\omega^{2}r}\right] \, .
\end{equation}
The Dirac-Coulomb Green's function $\tilde{G}$ in the presence of the magnetic field is determined by
\begin{equation}
\label{eq:Gfunc_magn}
\tilde{G}(\omega) = \sum_{n} 
\frac{|\tilde{n}\ra \la \tilde{n}|}
     {\omega-\tilde{\veps}_n +  i\tilde{\eta}_n0} \, ,
\end{equation}
where $\tilde{\eta}_n=\tilde{\veps}_n-\tilde{\veps}_{\rm F}$ and $\tilde{\veps}_{\rm F}$ is the Fermi energy which is chosen to be higher than the one-electron closed-shell energies and lower than the valence-electron energy $\tilde{\veps}_a$. In Eq.~(\ref{eq:g_general}) and throughout the paper, a summation over repeated indices is conventionally implied. Using the extended-nucleus potential $V_{\rm nucl}(r)$ as $V(r)$ in Eq.~(\ref{eq:Dirac}) allows one to partially take into account the nuclear size correction to the nuclear recoil effect on the $g$ factor~\cite{Shabaev:2001:052104}. In order to partially take into consideration the interelectronic-interaction effects already within the initial approximation, one can replace $V(r)$ with the effective potential $V_{\rm eff}(r)=V_{\rm nucl}(r)+V_{\rm scr}(r)$, where $V_{\rm scr}(r)$ is some local screening potential.

The fully relativistic expression (\ref{eq:g_general}) yields the nuclear recoil contribution to the bound-electron $g$ factor within the independent-electron approximation, i.e., to zeroth order in $1/Z$. At the same time, it allows one to deduce effective operators which can be employed in calculating the nuclear recoil effect within the lowest-order relativistic (Breit) approximation to all orders in $1/Z$ \cite{Shabaev:2017:263001}. These relativistic operators have the form
\begin{align} 
\label{eq:H_M_magn}
H_M^{\rm magn} &= -\mu_0 {\mathcal H} \frac{m}{M}
  \sum_{ij}
  \Big\{
      [\bfr_i\times \bfp_j]_z 
    - [\bfr_i\times \bfD_j(0)]_z
  \Big\} \, , \\
\label{eq:H_M}  
H_M &= \frac{1}{2M}
   \sum_{ij}
   \Big\{  
      \bfp_i \cdot \bfp_j
    - 2 \bfp_i \cdot \bfD_j(0)
    \Big\} \, ,
\end{align}
where
\begin{align}
\label{eq:D0}
\bfD(0) = \frac{\alpha Z}{2r} \left[ \balpha + \frac{(\balpha\cdot\bfr)}{r^2}\,\bfr \right] \, 
\end{align}
represents the zero-energy-transfer limit $\omega\to 0$ of the vector $\bfD(\omega)$. The first term in the curly braces in Eq.~(\ref{eq:H_M_magn}) corresponds to the nonrelativistic limit of the operator $H_M^{\rm magn}$ and was derived for the first time by Phillips~\cite{Phillips:1949:1803}. The mass operator~(\ref{eq:H_M}) describes also the nuclear recoil effect on the binding energies in the absence of the magnetic field \cite{Shabaev:1998:59}. In order to obtain the corresponding contribution to the atomic $g$ factor, one has to combine it with the operator of the interaction with the magnetic field,
\begin{align} 
\label{eq:H_magn}
H_{\rm magn} = \mu_0\mathcal{H} m \sum_{j} [ \bfr_j \times \balpha_j ]_z \, .
\end{align}
We note that the single-particle negative-continuum excitations are to be properly taken into account in the nuclear recoil contributions involving the operators $H_M$ and $H_{\rm magn}$ (see, e.g., Refs.~\cite{Glazov:2004:062104, Tupitsyn:2005:062503}). In what follows, we will refer to the Breit-approximation part of the nuclear recoil contribution which can be obtained with the aid of the effective operators as the low-order~($\rm L$) part as well.

The most straightforward way to construct the $1/Z$ perturbation theory for the nuclear recoil contribution to the $g$ factor within the Breit approximation is based on the Dirac-Coulomb-Breit (DCB) Hamiltonian for an ion which is placed in the magnetic field,
\begin{align}
\label{eq:H_DCB_magn}
\tilde{H}_{\rm DCB} &= 
  \tilde{\Lambda}^{(+)} \left[ 
    \tilde{H}_0 + H_{\rm int}
  \right] \tilde{\Lambda}^{(+)}
\, ,   \\ 
\label{eq:H0_magn}
\tilde{H}_0 &= \sum_j \tilde{h}_{\rm D}(j) \, , \\
\label{eq:int}
H_{\rm int} &= \alpha \sum_{i<j} 
     \left[ 
       \frac{1}{r_{ij}} - \frac{ {\balpha}_i \cdot {\balpha}_j }{ r_{ij} }
      - \frac{1}{2} ( \balpha_i \cdot \bnabla_i ) ( \balpha_j \cdot \bnabla_j ) r_{ij} 
     \right]
\, .
\end{align}
If a screening potential is included into the Dirac Hamiltonian, the counterterm $\delta H_{\rm int} = -\sum_j V_{\rm scr}(r_j)$ has to be added to $H_{\rm int}$. The projection operator $\tilde{\Lambda}^{(+)}$ in Eq.~(\ref{eq:H_DCB_magn}) is constructed as a product of the one-electron positive-energy-states projectors corresponding to the Hamiltonian $\tilde{h}_{\rm D}$ in Eq.~(\ref{eq:Dirac_magn}). The conventional DCB Hamiltonian (in the absence of the magnetic field) can be written as
\begin{align}
\label{eq:H_DCB}
H_{\rm DCB} &= 
  \Lambda^{(+)} \left[ 
    H_0 + H_{\rm int}
  \right] \Lambda^{(+)}
\, , \\ 
\label{eq:H0}
H_0 &= \sum_j h_{\rm D}(j) \, ,
\end{align}
where the projector $\Lambda^{(+)}$ corresponds to the one-electron Dirac operator $h_{\rm D}$ in Eq.~(\ref{eq:Dirac}). We stress that the Hamiltonians $\tilde{H}_{\rm DCB}$ and $H_{\rm DCB}$ differ not only due to the term describing the interaction with the magnetic field, $\tilde{H}_0 = H_0+H_{\rm magn}$, but also in the definition of the projection operators in Eqs.~(\ref{eq:H_DCB_magn}) and (\ref{eq:H_DCB}).

To zeroth order in $1/Z$, the Breit-approximation recoil contribution to the bound-electron $g$ factor can be expressed by employing the operators (\ref{eq:H_M_magn}), (\ref{eq:H_M}), and (\ref{eq:H_magn}) as follows:
\begin{align}
\label{eq:breit0}
\Delta g_{\rm L}^{(0)} &= 
\left.
\frac{1}{\mu_0 m_a}\, \frac{\partial}{\partial{\mathcal H}}
\la \tilde{A} | \left[ H_M^{\rm magn} + H_M \right] | \tilde{A} \ra
\right|_{\mathcal H=0}     \nonumber \\
&=
\frac{1}{\mu_0 {\mathcal H} m_a}
\left\{ 
\la A | H_M^{\rm magn} | A \ra 
+
2\, 
{\sum_{N}}'\,\frac{\la A | H_M | N \ra \la N | H_{\rm magn} | A \ra} {E_A - E_N}
\right\} \, .
\end{align}
Here, $|\tilde{A}\rangle$ and $|A\rangle$ denote the many-electron wave functions of the state of interest which are evaluated in the presence of the magnetic field,  $\tilde{H}_0|\tilde{A}\rangle = \tilde{E}_A|\tilde{A}\rangle$, and in its absence, $H_0|A\rangle = E_A|A\rangle$, respectively. They are constructed from the eigenfunctions of the corresponding one-electron Dirac Hamiltonians $\tilde{h}_{\rm D}$ and $h_{\rm D}$. The many-electron energies $\tilde{E}_A$ and $E_A$ are equal to the sum of the one-electron Dirac energies. For instance, $E_A=\veps_a + \sum_c \veps_c $, where the summation over all of the closed-shell electronic states~$|c\rangle$ is introduced. In turn, the summation over $|N\rangle$ in Eq.~(\ref{eq:breit0}) is performed over the complete many-electron spectrum of the Hamiltonian $H_0$, i.e., $H_0|N\rangle = E_N|N\rangle$. The one-electron negative-energy excitations from the state $|A\rangle$ are included into the summation. The prime by the summation symbol in Eq.~(\ref{eq:breit0}) and in what follows indicates that the terms with vanishing denominators are omitted from the summation. The contribution $\Delta g_{\rm L}^{(0)}$ is conveniently divided into a one-electron (the terms with $i=j$ in $H_M^{\rm magn}$ and $H_M$) and two-electron (the terms with $i\neq j$) parts, 
\begin{align}
\label{eq:breit0_terms}
\Delta g_{\rm L}^{(0)}=\Delta g_{\rm L\text{-}1el}^{(0)}+\Delta g_{\rm L\text{-}2el}^{(0)} \, .  
\end{align}
In the case of the ground $(1s)^2 2s$ state of Li-like ions, the two-electron part vanishes to zeroth order in $1/Z$ while the one-electron part is of pure relativistic origin~\cite{Shabaev:2017:263001, Shabaev:2018:032512}. For B-like ions, the nuclear recoil contributions $\Delta g_{\rm L\text{-}1el}^{(0)}$ and $\Delta g_{\rm L\text{-}2el}^{(0)}$ possess a nonzero nonrelativistic limit, and the corresponding $\alpha Z$-expansions start, therefore, from the $(m/M)(\alpha Z)^0$ term~\cite{Aleksandrov:2018:062521}. We note that the contribution $\Delta g_{\rm L}^{(0)}$ can be obtained directly from Eq.~(\ref{eq:g_general}) by considering it within the Breit approximation.

The first order of perturbation theory in $H_{\rm int}$ leads to the following correction to the nuclear recoil contribution:
\begin{align}
\label{eq:breit1}
\Delta g_{\rm L}^{(1)} &= 
\frac{2}{\mu_0 m_a}\,\frac{\partial}{\partial {\mathcal H}}
 \left.
   \psum_{N}^{(+)} \, 
 \frac{\la\tilde{A} | \left[ H_M^{\rm magn} + H_M \right] | \tilde{N}\ra
       \la\tilde{N} | H_{\rm int} | \tilde A \ra} {\tilde E_A - \tilde E_N}
    \right|_{\mathcal H=0}
\,,
\end{align}
where the plus sign over the sum indicates that the intermediate states $|\tilde{N}\rangle$ are constructed only from the positive-energy eigenfunctions of the Dirac Hamiltonian $\tilde{h}_{\rm D}$, i.e., $\big[\tilde{\Lambda}^{(+)}\tilde{H}_0\tilde{\Lambda}^{(+)}\big]|\tilde{N}\rangle = \tilde{E}_N|\tilde{N}\rangle$. The part of the expression~(\ref{eq:breit1}) corresponding to the operator $H_M^{\rm magn}$ can be readily rewritten in the explicit form. Indeed, we obtain 
\begin{align}
\label{eq:breit1_magn}
\Delta g_{\rm L\text{-}magn}^{(1)} &= 
   \frac{2}{\mu_0 {\mathcal H} m_a}\,
   \psum_{N}^{(+)} \, 
   \frac{\la A | H^\text{magn}_M | N \ra \la N | H_\text{int} | A \ra} {E_A - E_N} \, ,
\end{align}
where $\big[\Lambda^{(+)}H_0\Lambda^{(+)}\big]|N\rangle = E_N|N\rangle$. The explicit form of the second part of Eq.~(\ref{eq:breit1}) is rather cumbersome, so we do not display it here. The contribution $\Delta g_{\rm L}^{(1)}$ is evaluated in the present study. The higher-order interelectronic-interaction corrections to the nuclear recoil effect on the $g$ factor can be obtained by employing, e.g., the configuration-interaction method or the recursive representation of perturbation theory~\cite{Glazov:2017:46}. However, this is beyond the scope of the present investigation.

Let us now return to the initial fully relativistic formula (\ref{eq:g_general}) and discuss the higher-order~($\rm H$) in $\alpha Z$ contribution which is not covered by the effective operators $H_M^{\rm magn}$ and $H_M$. The derivation of this contribution demands the bound-state QED beyond the Breit approximation be applied. For this reason, we will also refer to this term as the QED term. The one-electron part of the higher-order nuclear recoil contribution corresponding to a valence-electron state $|a\rangle$ has the form
\begin{align}
\label{eq:g_H}
\Delta g_{\rm H\text{-}1el}^{(0)}
=&\, \frac{1}{m_a} \frac{m}{M} \frac{i}{2\pi}
    \int_{-\infty}^{\infty} \! d\omega \,
    \Bigl\{ \la \delta a | B^k_{-}(\omega) G(\omega+\veps_a) B^k_{+}(\omega) | a \ra          
                 + \la a | B^k_{-}(\omega) G(\omega+\veps_a) B^k_{+}(\omega) | \delta a \ra   \nonumber\\ 
& \qquad\qquad
+ \la a | B^k_{-}(\omega) G(\omega+\veps_a) 
          \Big( [ \bfr\times\balpha ]_z - \langle a | [\bfr\times\balpha]_z | a \rangle \Big) 
                          G(\omega+\veps_a) B^k_{+}(\omega) | a \ra \Bigr\} \, ,  
\end{align}
where $|\delta a\ra = \sum_n^{\veps_n\ne \veps_a} | n \ra \la n | [\bfr\times\balpha]_z | a \ra (\veps_a-\veps_n)^{-1}$ is the correction to the wave function due to the external magnetic field, $B^k_{\pm}(\omega)=D^k(\omega)\pm [p^k,V]/(\omega+i0)$, $[A,B]=AB-BA$, and the conventional Dirac-Coulomb Green's function (in the absence of $\boldsymbol{\mathcal H}$) can be represented as
\begin{align}
\label{eq:Gfunc}
G(\omega) = \sum_{n} 
\frac{|n\ra \la n|}
     {\omega-\veps_n (1-i0)} \, .
\end{align}
The explicit formula for the two-electron part $\Delta g^{(0)}_{\rm H\text{-}2el}$ of the higher-order nuclear recoil contribution
to the $g$ factor of the state under consideration is rather lengthy. Therefore, we present a more compact expression for the fully relativistic two-electron contribution $\Delta g^{(0)}_{\rm 2el}$ which arises from Eq.~(\ref{eq:g_general}) and includes the corresponding part of the low-order term. Within the independent-electron approximation, the two-electron recoil contribution valid to all orders in $\alpha Z$ can be written as follows:
\begin{align}
\label{eq:QED_2el}
\Delta g^{(0)}_{\rm 2el} & 
   = \Delta g_{\rm L\text{-}2el}^{(0)} + 
     \Delta g^{(0)}_{\rm H\text{-}2el} 
\nonumber \\ 
& = \frac{1}{m_a} \frac{m}{M} \sum_c \, 
  \Big\{ \epsilon_{3kl} \left( \la a |r^k| c \ra \la c |[p^l-D^l(\Delta)]| a \ra 
                         + \la a |[p^l-D^l(\Delta)]| c \ra \la c |r^k| a \ra \right)
\nonumber\\
&\quad - \la \delta a |[p^k-D^k(\Delta)]| c \ra \la c |[p^k-D^k(\Delta)]| a \ra
    - \la a |[p^k-D^k(\Delta)]| \delta c \ra \la c |[p^k-D^k(\Delta)]| a \ra
\nonumber\\
&\quad - \la a |[p^k-D^k(\Delta)]| c \ra \la \delta c |[p^k-D^k(\Delta)]| a \ra
    - \la a |[p^k-D^k(\Delta)]| c \ra \la c |[p^k-D^k(\Delta)]| \delta a \ra
\nonumber\\
&\quad + \left(
      \la a |\frac{dD^k(\omega)}{d\omega}\Bigr|_{\omega=\Delta}| c \ra \la c |[p^k-D^k(\Delta)]| a \ra 
    + \la a |[p^k-D^k(\Delta)]| c \ra \la c |\frac{dD^k(\omega)}{d\omega}\Bigr|_{\omega=\Delta}| a \ra 
    \right)
\nonumber\\
&\quad \times 
      \big( 
      \langle a | [\bfr\times\balpha]_z | a \rangle - \langle c | [\bfr\times\balpha]_z | c \rangle 
      \big)
  \Big\}
\, , 
\end{align}
where $\epsilon_{ikl}$ is the Levi-Civita symbol, the summation runs over the closed-shell electronic states $|c\rangle$, $\Delta=\veps_a-\veps_c$, and $|\delta c\ra=\sum_n^{\veps_n\ne \veps_c}|n\ra\la n|[\bfr\times\balpha]_z|c\ra (\veps_c-\veps_n)^{-1}$. The expression (\ref{eq:QED_2el}) is reduced to the low-order part $\Delta g_{\rm L\text{-}2el}^{(0)}$ if one sets $\Delta=0$ (in this limit the terms with  $d\bfD/d\omega$ vanish) and discards the $\bfD\cdot\bfD$ products. The higher-order part 
$\Delta g^{(0)}_{\rm H\text{-}2el}$ 
corresponds to the remainder.


\section{Results and discussion}

\input{table_1el_Low_2p3_Z18-Z92.tex}

In this section, we present the results of our evaluation of the nuclear recoil effect on the $g$ factor of the ${}^2P_{3/2}$ state in B-like ions. The initial approximation is determined by the Dirac Hamiltonian $h_{\rm D}$ with the spherically symmetric potential $V(r)$. The calculations are performed with the Coulomb potential as $V(r)$ in Eq.~(\ref{eq:Dirac}) and with several effective potentials. Namely, we use the core-Hartree (CH), Perdew-Zunger (PZ)~\cite{pot:PZ}, Kohn-Sham (KS)~\cite{pot:KS}, and local Dirac-Fock (LDF)~\cite{Shabaev:2005:062105} potentials. These coincide with the effective potentials used in Ref.~\cite{Aleksandrov:2018:062521}. Computations with an effective potential allow one to partially incorporate the higher-order interelectronic-interaction corrections. Applying various initial approximations provides an estimate of the uncalculated higher-order contributions. In order to describe the nuclear charge distribution, we use the Fermi model with the nuclear charge radii taken from Ref.~\cite{Angeli:2013:69}. The summation over the intermediate electronic states is carried out by employing finite-basis sets constructed from $B$ splines \cite{Sapirstein:1996:5213} within the dual-kinetic-balance approach \cite{splines:DKB}.

\input{table_B_2el_2p3_Z18-Z92.tex}

For presentation of the low-order (Breit-approximation) contribution $\Delta g^{(0)}_{\rm L}$ to the $g$ factor of B-like ions obtained within the independent-electron approximation, we introduce the function~$A(\alpha Z)$ defined according to
\begin{align}
\label{A_def}
  \Delta g^{(0)}_{\rm L} &= \frac{m}{M} A(\alpha Z) \, .
\end{align}
The one-electron part $\Delta g^{(0)}_{\rm L\text{-}1el}$ of the low-order contribution is given in Table~\ref{tab:1el_low} in terms of the function $A(\alpha Z)$. The results obtained for the pure Coulomb potential of the nucleus are presented in the second column while the values calculated with the effective potentials are shown in the subsequent columns. The two-electron nuclear recoil contribution $\Delta g^{(0)}_{\rm 2el}$ to the $^2P_{3/2}$-state $g$ factor of B-like ions to zeroth order in $1/Z$ is presented in terms of the function $A(\alpha Z)$ in Table~\ref{tab:2el}. For each $Z$, the low-order two-electron contribution $\Delta g_{\rm L\text{-}2el}^{(0)}$ and the higher-order correction $\Delta g^{(0)}_{\rm H\text{-}2el}$ are shown separately. It is seen that the latter grows rapidly and reaches approximately $5\%$ for B-like uranium.

\input{table_1el_QED_2p3_Z18-Z92.tex}

The higher-order (QED) correction $\Delta g^{(0)}_{\rm H\text{-}1el}$ to the one-electron part of the nuclear recoil contribution $\Delta g^{(0)}_{\rm L\text{-}1el}$ is conveniently represented via the function $P(\alpha Z)$,
\begin{align}  
\label{P_def}
\Delta g^{(0)}_{\rm H\text{-}1el} 
&= \frac{m}{M} \frac{(\alpha Z)^3}{8} P(\alpha Z) \, .
\end{align}
Recently, it was shown that for $p$ states this contribution behaves as $(m/M)(\alpha Z)^3$ \cite{Malyshev:2019:G_rec_H:preprint}, in contrast to $s$ states, where it exhibits the $(m/M)(\alpha Z)^5$ behavior. The results for the valence $2p_{3/2}$ state obtained for the Coulomb and four effective potentials are presented in Table~\ref{tab:1el_QED} in terms of the function~$P(\alpha Z)$. By comparing Tables~\ref{tab:1el_low} and \ref{tab:1el_QED} and taking into account Eqs.~(\ref{A_def}) and (\ref{P_def}), one observes that the QED correction to the one-electron nuclear recoil contribution grows monotonically and reaches approximately $3\%$ for $Z=92$.

\input{table_IntEl1_2p3_Z18-Z92.tex}

\input{table_01orders_Breit_2p3_Z18-Z92.tex}

The first-order (in $1/Z$) interelectronic-interaction correction to the nuclear recoil effect on the $^2 P_{3/2}$-state $g$ factor is presented in Table~\ref{tab:IntEl1} in terms of the function $B(\alpha Z)$,
\begin{align}  
\label{B_def}
\Delta g^{(1)}_{\rm L} &= \frac{m}{M} \frac{B(\alpha Z)}{Z} \, .
\end{align}
It is seen that the individual terms shown in Tables~\ref{tab:1el_low}, \ref{tab:2el}, and \ref{tab:IntEl1} may significantly vary from one initial approximation to another. However, the complete Breit-approximation value, which is a sum of the low-order contributions of zeroth and first orders in $1/Z$,
\begin{align}
\label{eq:breit01}
\Delta g_{\rm Breit} = \Delta g^{(0)}_{\rm L} + \Delta g^{(1)}_{\rm L} \, ,
\end{align}
is much more stable. This is demonstrated for $Z=18$, $Z=50$, and $Z=92$ in Table~\ref{tab:IntEl01}, where the zeroth-order contribution $A=A_{\rm 1el}+A_{\rm 2el}$ and the first-order correction $B/Z$ are shown together with their sum. The data are taken from Tables~\ref{tab:1el_low}, \ref{tab:2el} (the rows labeled with $\rm L$), and \ref{tab:IntEl1}.

\input{table_B_01_total_2p3_Z18-Z92.tex}

\begin{figure}
\begin{center}
\includegraphics[width=0.65\linewidth]{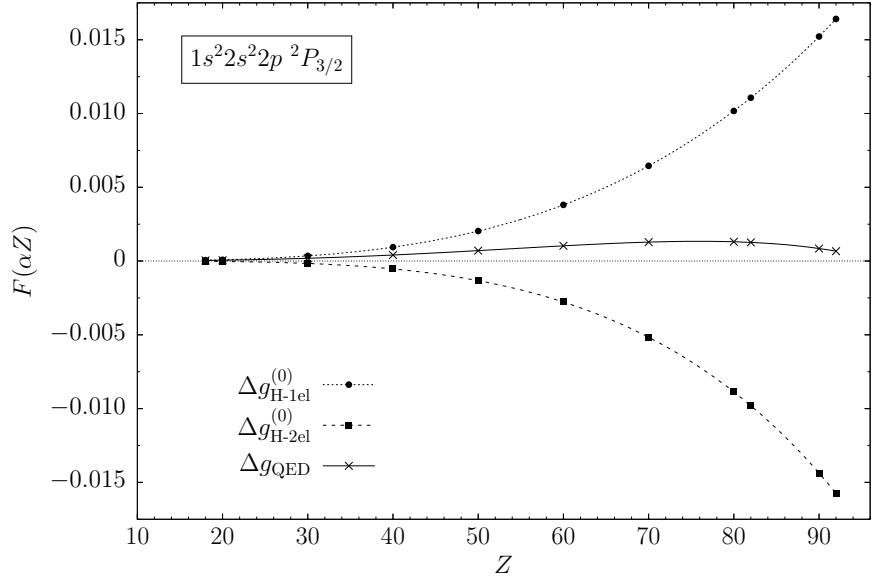}
\caption{\label{fig:qed_2p3}
Higher-order (QED) recoil contribution to the $^2P_{3/2}$-state $g$ factor of B-like ions. The one-electron part $\Delta g_{\rm H\text{-}1el}^{(0)}$ and two-electron part $\Delta g_{\rm H\text{-}2el}^{(0)}$ constituting the total QED contribution $\Delta g_{\rm QED}=\Delta g_{\rm H\text{-}1el}^{(0)}+\Delta g_{\rm H\text{-}2el}^{(0)}$ are shown separately. The results are expressed in terms of the function $F(\alpha Z)$ defined by Eq.~(\ref{F_def}).
}
\end{center}
\end{figure}

\begin{figure}
\begin{center}
\includegraphics[width=0.65\linewidth]{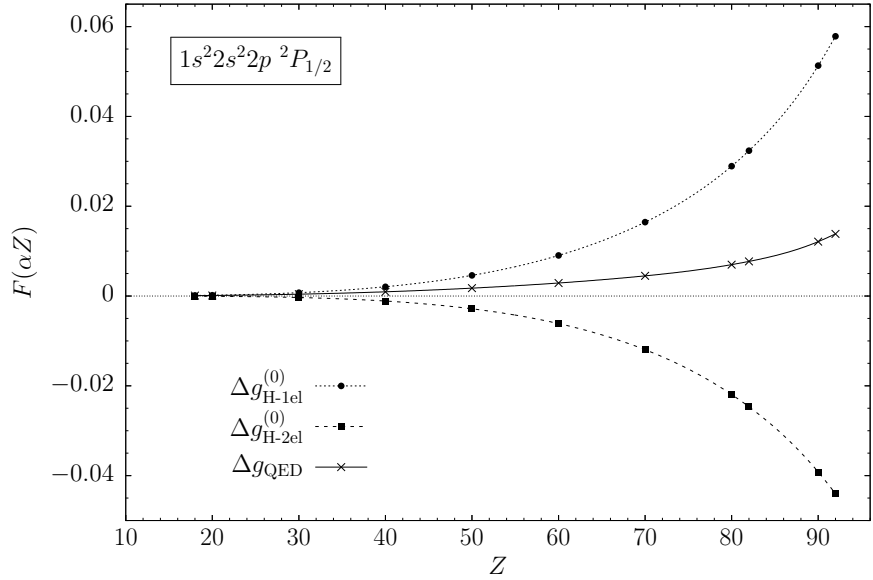}
\caption{\label{fig:qed_2p1}
Higher-order (QED) recoil contribution to the $^2P_{1/2}$-state $g$ factor of B-like ions. The one-electron part $\Delta g_{\rm H\text{-}1el}^{(0)}$ and two-electron part $\Delta g_{\rm H\text{-}2el}^{(0)}$ constituting the total QED contribution $\Delta g_{\rm QED}=\Delta g_{\rm H\text{-}1el}^{(0)}+\Delta g_{\rm H\text{-}2el}^{(0)}$ are shown separately. The results are expressed in terms of the function $F(\alpha Z)$ defined by Eq.~(\ref{F_def}).
}
\end{center}
\end{figure}

The total value of the nuclear recoil contribution to the $g$ factor of the $^2 P_{3/2}$ state is presented in Table~\ref{tab:total}. It is obtained by summing all of the corrections discussed above. The results calculated with the LDF potential are used as the final values. In Table~\ref{tab:total} we represent the total recoil contribution as a sum of the low-order (Breit) part, which is given by Eq.~(\ref{eq:breit01}), and the higher-order (QED) part,
\begin{align}
\label{eq:g_tot}
\Delta g_{\rm rec} = \Delta g_{\rm Breit} + \Delta g_{\rm QED} \, .
\end{align}
The QED contribution incorporates the data given in Tables~\ref{tab:1el_QED} and \ref{tab:2el} (the rows $\rm H$),
\begin{align}  
\label{eq:g_QED}
\Delta g_{\rm QED} = \Delta g^{(0)}_{\rm H\text{-}1el} + \Delta g^{(0)}_{\rm H\text{-}2el} \, .   
\end{align}
All the results in Table~\ref{tab:total} are shown in terms of the mass-ratio-independent function $F(\alpha Z)$,
\begin{align}  
\label{F_def}
  \Delta g &= \frac{m}{M} F(\alpha Z) \, ,     
\end{align}
so that
\begin{align}  
\label{F_def_sum}
F_{\rm rec}(\alpha Z) = F_{\rm Breit}(\alpha Z) + F_{\rm QED}(\alpha Z) \, .  
\end{align}
We see that the QED contribution found is rather small compared to the Breit-approximation part of the nuclear recoil effect over the whole $Z$ range considered here. Moreover, its absolute value starts to decrease once $Z\gtrsim 80$. This results from a strong cancellation appearing between the one- and two-electron higher-order contributions, which is demonstrated in Fig.~\ref{fig:qed_2p3}, where these contributions are plotted in terms of $F(\alpha Z)$ together with their sum. A similar situation takes place for the ground~$^2 P_{1/2}$ state of B-like ions which was investigated in Ref.~\cite{Aleksandrov:2018:062521}. The one-electron and two-electron QED contributions along with the total QED part of the nuclear recoil effect on the $^2 P_{1/2}$-state $g$ factor are shown in Fig.~\ref{fig:qed_2p1} in terms of the function $F(\alpha Z)$. It is seen that the individual terms considerably cancel each other when summed to receive the total QED value. Nevertheless, in contrast to the $^2 P_{3/2}$ state, for the ground state the total QED contribution is a monotonic function of $Z$.

There are several sources of the theoretical uncertainties given in Table~\ref{tab:total} in the parentheses. First of all, we estimate the uncertainty due to the incomplete treatment of the interelectronic-interaction correction to the nuclear recoil effect. For the Breit part, in addition to the scatter of the results obtained in the calculations with the different effective potentials, we include into the uncertainty the term $(\Delta g_{\rm L}^{(1)}/\Delta g_{\rm L}^{(0)})\cdot\Delta g_{\rm L}^{(1)}$ evaluated for the Coulomb potential. This is done in order not to underestimate the higher-order (in $1/Z$) two-electron recoil contribution to the bound-electron $g$ factor (see the related discussion in Refs.~\cite{Aleksandrov:2018:062521, Glazov:2018:457}). For the QED part, the spread of the results for the different effective potentials is supplemented by a rather conservative estimate $(2/Z)\cdot\Delta g_{\rm H\text{-}1el}$. We employ here the one-electron higher-order contribution since the heavy cancellation of the one- and two-electron terms revealed in zeroth order in $1/Z$ could not take place in higher orders. Second, we take into account the uncertainty due to the approximate treatment of the nuclear size correction. As was noted in Ref.~\cite{Shabaev:2001:052104}, using the extended-nucleus potential in the initial approximation provides only a partial description of this effect. In order to estimate the corresponding uncertainty, we assume that in the case of the bound-electron $g$ factor the relative uncertainty due to this approximation is the same as that found for the binding energy \cite{Aleksandrov:2015:144004} (see also the related discussion, e.g., in Refs.~\cite{Shabaev:2018:032512, Aleksandrov:2018:062521}). The uncertainties indicated above are then combined by evaluating their root sum square.

\input{table_isotopes_2p3_Z18-Z92.tex}

Finally, in Table~\ref{tab:isotopes} we present the nuclear recoil contribution to the $g$ factor of the $^2P_{3/2}$ state in selected B-like ions in the range $Z=18$--$92$. The nuclear masses for isotopes are taken from the Ame2012 compilation~\cite{Wang:2012:1603} in accordance with Ref.~\cite{Yerokhin:2015:033103}. The nuclear recoil effect on the $^2P_{3/2}$-state $g$ factor of B-like argon ($Z=18$) was studied in Ref.~\cite{Shchepetnov:2015:012001}. The calculations performed therein did not include the one-electron higher-order contribution $\Delta g_{\rm H\text{-}1el}$ whereas the $1/Z$ interelectronic-interaction correction was evaluated only for the nonrelativistic part of the operator $H^{\rm magn}_{\rm M}$. The new value of the nuclear recoil contribution for B-like argon is found to be in agreement with the previous one. 

The new generation of experiments measuring the bound-electron $g$ factor of B-like ions is aimed at reaching an accuracy of $10^{-9}$ and better~\cite{Arapoglou:2019:253001}. From Table~\ref{tab:isotopes}, one can conclude that the nuclear recoil correction to the $^2P_{3/2}$-state $g$ factor represents a significant contribution to the total value. For high-$Z$ B-like ions, the QED part exceeds the current theoretical uncertainty, see Table~\ref{tab:total}. For further improvement of the theoretical accuracy, one has to perform the calculations of the second- and higher-order interelectronic-interaction corrections to the Breit part as well as the $1/Z$ correction to the QED part. We will address these problems in our future investigations.

\section{Conclusion}

In the present study, we evaluated the nuclear recoil effect on the $g$ factor of the $^2 P_{3/2}$ state in B-like ions in the range $Z=18$--$92$. The leading one-electron and two-electron recoil contributions are calculated within the rigorous QED formalism to all orders in the parameter $\alpha Z$. The first-order interelectronic-interaction correction to the nuclear recoil effect is obtained within the Breit approximation by employing the effective relativistic operators. The higher-order (in $1/Z$) contributions are partially taken into account by means of the effective potential. As a result, the most precise theoretical predictions for the nuclear recoil contribution to the $^2 P_{3/2}$-state $g$ factor are presented. 

\section{Acknowledgments}

This work was supported by the Russian Science Foundation (Grant No. 17-12-01097).




\end{document}

%% file: table_1el_Low_2p3_Z18-Z92.tex
\begin{table}[t]
\centering

\renewcommand{\arraystretch}{1.1}

\caption{\label{tab:1el_low}
Low-order one-electron recoil contribution 
$\Delta g_{\rm L\text{-}1el}^{(0)}$ 
to
the $g$ factor of the $2p_{3/2}$ state. The results are presented in terms of
the function $A(\alpha Z)$ defined by Eq.~(\ref{A_def}). 
Coul, CH, PZ, KS, and LDF
refer to the calculations with the Coulomb and various
screening potentials (see the text).
}

\begin{tabular}{l@{\quad\,}
                S[table-format=-1.6,group-separator=]@{\quad\,}
                S[table-format=-1.6,group-separator=]@{\quad\,}
                S[table-format=-1.6,group-separator=]@{\quad\,}
                S[table-format=-1.6,group-separator=]@{\quad\,}
                S[table-format=-1.6,group-separator=]
               }
               
\hline
\hline

       
 \multirow{2}{*}{$Z$} &
 \multicolumn{5}{c}{$A_{\rm1el}(\alpha Z)$} \\
 &  {Coul} 
 &  {CH  }
 &  {PZ  }
 &  {KS  }
 &  {LDF }  \\
        
\hline        

 18  &    -0.662641  &    -0.663435  &    -0.663311  &    -0.663310  &    -0.663384     \\ 

 20  &    -0.661697  &    -0.662583  &    -0.662444  &    -0.662445  &    -0.662525     \\ 

 30  &    -0.655489  &    -0.656845  &    -0.656628  &    -0.656637  &    -0.656752     \\ 

 40  &    -0.646801  &    -0.648643  &    -0.648349  &    -0.648369  &    -0.648520     \\ 

 50  &    -0.635641  &    -0.637990  &    -0.637621  &    -0.637655  &    -0.637842     \\ 

 60  &    -0.622014  &    -0.624900  &    -0.624457  &    -0.624512  &    -0.624733     \\ 

 70  &    -0.605929  &    -0.609395  &    -0.608877  &    -0.608961  &    -0.609212     \\ 

 80  &    -0.587398  &    -0.591498  &    -0.590907  &    -0.591032  &    -0.591303     \\ 

 82  &    -0.583399  &    -0.587635  &    -0.587028  &    -0.587164  &    -0.587438     \\ 

 90  &    -0.566432  &    -0.571244  &    -0.570578  &    -0.570763  &    -0.571038     \\ 

 92  &    -0.561948  &    -0.566914  &    -0.566232  &    -0.566432  &    -0.566706     \\

\hline
\hline

\end{tabular}%

\end{table}

%% file: table_B_2el_2p3_Z18-Z92.tex
\begin{table}[t]
\centering

\renewcommand{\arraystretch}{1.0}

\caption{\label{tab:2el}
Two-electron nuclear-recoil contribution 
$\Delta g_{\rm 2el}^{(0)}$ 
to
the $^2P_{3/2}$-state $g$ factor of B-like ions to zeroth order in $1/Z$. 
The low-order part 
$\Delta g_{\rm L\text{-}2el}^{(0)}$ 
and the higher-order part 
$\Delta g_{\rm H\text{-}2el}^{(0)}$ 
are given
separately in the lines labeled with $\rm L$ and $\rm H$, respectively. The results
are expressed in terms of the function $A(\alpha Z)$ defined by Eq.~(\ref{A_def}).
}

\begin{tabular}{l@{\quad}
                c@{\quad}
                S[table-format=-1.6,group-separator=]@{\quad\,}
                S[table-format=-1.6,group-separator=]@{\quad\,}
                S[table-format=-1.6,group-separator=]@{\quad\,}
                S[table-format=-1.6,group-separator=]@{\quad\,}
                S[table-format=-1.6,group-separator=]                 
               }
               
\hline
\hline

       
 \multirow{2}{*}{$Z$}  &
 \multirow{2}{*}{Part} &
 \multicolumn{5}{c}{$A_{\rm 2el}(\alpha Z)$} \\
 &
 &  {Coul} 
 &  {CH  }
 &  {PZ  }
 &  {KS  }
 &  {LDF }  \\     
       
\hline        

  \multirow{2}{*}{18}  &  L  &     0.278294  &     0.302879  &     0.303671  &     0.307182  &     0.303870    \\ 
                       &  H  &    -0.000026  &    -0.000018  &    -0.000019  &    -0.000018  &    -0.000018    \\[1.5mm] 

  \multirow{2}{*}{20}  &  L  &     0.278503  &     0.300270  &     0.300992  &     0.304014  &     0.301128    \\ 
                       &  H  &    -0.000039  &    -0.000028  &    -0.000030  &    -0.000028  &    -0.000029    \\[1.5mm] 

  \multirow{2}{*}{30}  &  L  &     0.280022  &     0.294232  &     0.294709  &     0.296513  &     0.294702    \\ 
                       &  H  &    -0.000197  &    -0.000160  &    -0.000164  &    -0.000160  &    -0.000162    \\[1.5mm] 

  \multirow{2}{*}{40}  &  L  &     0.282575  &     0.293564  &     0.293909  &     0.295237  &     0.293858    \\ 
                       &  H  &    -0.000616  &    -0.000528  &    -0.000538  &    -0.000530  &    -0.000533    \\[1.5mm] 

  \multirow{2}{*}{50}  &  L  &     0.286619  &     0.295912  &     0.296182  &     0.297274  &     0.296116    \\ 
                       &  H  &    -0.001486  &    -0.001316  &    -0.001336  &    -0.001320  &    -0.001326    \\[1.5mm] 

  \multirow{2}{*}{60}  &  L  &     0.292789  &     0.301078  &     0.301312  &     0.302277  &     0.301235    \\ 
                       &  H  &    -0.003050  &    -0.002756  &    -0.002791  &    -0.002764  &    -0.002774    \\[1.5mm] 

  \multirow{2}{*}{70}  &  L  &     0.301922  &     0.309554  &     0.309782  &     0.310681  &     0.309691    \\ 
                       &  H  &    -0.005605  &    -0.005137  &    -0.005195  &    -0.005153  &    -0.005166    \\[1.5mm] 

  \multirow{2}{*}{80}  &  L  &     0.315116  &     0.322255  &     0.322507  &     0.323375  &     0.322391    \\ 
                       &  H  &    -0.009534  &    -0.008825  &    -0.008916  &    -0.008854  &    -0.008868    \\[1.5mm] 

  \multirow{2}{*}{82}  &  L  &     0.318355  &     0.325403  &     0.325663  &     0.326529  &     0.325540    \\ 
                       &  H  &    -0.010526  &    -0.009759  &    -0.009858  &    -0.009792  &    -0.009805    \\[1.5mm] 

  \multirow{2}{*}{90}  &  L  &     0.333669  &     0.340352  &     0.340654  &     0.341516  &     0.340499    \\ 
                       &  H  &    -0.015348  &    -0.014306  &    -0.014442  &    -0.014356  &    -0.014367    \\[1.5mm] 

  \multirow{2}{*}{92}  &  L  &     0.338148  &     0.344735  &     0.345051  &     0.345912  &     0.344886    \\ 
                       &  H  &    -0.016795  &    -0.015673  &    -0.015821  &    -0.015728  &    -0.015738    \\[1.5mm]

\hline
\hline

\end{tabular}%

\end{table}

%% file: table_1el_QED_2p3_Z18-Z92.tex
\begin{table}[t]
\centering

\renewcommand{\arraystretch}{1.1}

\caption{\label{tab:1el_QED}
Higher-order (QED) one-electron recoil contribution
$\Delta g_{\rm H\text{-}1el}^{(0)}$ 
to the $g$ factor of the $2p_{3/2}$ state. The results are expressed in
terms of the function $P(\alpha Z)$ defined by Eq.~(\ref{P_def}).
}

\begin{tabular}{l@{\quad\,}
                S[table-format=1.5,group-separator=]@{\quad\,}
                S[table-format=1.5,group-separator=]@{\quad\,}
                S[table-format=1.5,group-separator=]@{\quad\,}
                S[table-format=1.5,group-separator=]@{\quad\,}
                S[table-format=1.5,group-separator=]                 
               }
               
\hline
\hline

       
 \multirow{2}{*}{$Z$} &
 \multicolumn{5}{c}{$P(\alpha Z)$} \\
 &  {Coul} 
 &  {CH  }
 &  {PZ  }
 &  {KS  }
 &  {LDF }  \\       
        
\hline        

 18  &      0.26965  &      0.20958  &      0.21643  &      0.20984  &      0.21363     \\ 

 20  &      0.27637  &      0.22021  &      0.22667  &      0.22056  &      0.22403     \\ 

 30  &      0.30917  &      0.26450  &      0.26966  &      0.26490  &      0.26749     \\ 

 40  &      0.33996  &      0.30103  &      0.30546  &      0.30136  &      0.30354     \\ 

 50  &      0.36816  &      0.33285  &      0.33678  &      0.33309  &      0.33502     \\ 

 60  &      0.39352  &      0.36080  &      0.36438  &      0.36099  &      0.36271     \\ 

 70  &      0.41619  &      0.38543  &      0.38875  &      0.38560  &      0.38714     \\ 

 80  &      0.43681  &      0.40752  &      0.41065  &      0.40770  &      0.40908     \\ 

 82  &      0.44079  &      0.41173  &      0.41483  &      0.41192  &      0.41326     \\ 

 90  &      0.45666  &      0.42830  &      0.43133  &      0.42854  &      0.42975     \\ 

 92  &      0.46070  &      0.43246  &      0.43548  &      0.43271  &      0.43389     \\

\hline
\hline

\end{tabular}%

\end{table}

%% file: table_IntEl1_2p3_Z18-Z92.tex
\begin{table}[t]
\centering

\renewcommand{\arraystretch}{1.1}

\caption{\label{tab:IntEl1}
$1/Z$ interelectronic-interaction correction $\Delta g_{\rm L}^{(1)}$ to
 the nuclear-recoil effect on the $^2P_{3/2}$-state $g$ factor of B-like ions
 evaluated within the Breit approximation. The results are expressed
 in terms of the function $B(\alpha Z)$ defined by Eq.~(\ref{B_def}).
}

\begin{tabular}{l@{\quad\,}
                S[table-format=1.5,group-separator=]@{\quad\,}
                S[table-format=1.5,group-separator=]@{\quad\,}
                S[table-format=1.5,group-separator=]@{\quad\,}
                S[table-format=1.5,group-separator=]@{\quad\,}
                S[table-format=1.5,group-separator=]                 
               }
               
\hline
\hline

       
 \multirow{2}{*}{$Z$} &
 \multicolumn{5}{c}{$B(\alpha Z)$} \\
 &  {Coul} 
 &  {CH  }
 &  {PZ  }
 &  {KS  }
 &  {LDF }  \\       
        
\hline        

 18  &      0.91674  &      0.59705  &      0.57342  &      0.50157  &      0.57305     \\ 

 20  &      0.91361  &      0.59287  &      0.56908  &      0.50150  &      0.56983     \\ 

 30  &      0.89264  &      0.57019  &      0.54475  &      0.48712  &      0.55026     \\ 

 40  &      0.86203  &      0.54328  &      0.51416  &      0.45917  &      0.52443     \\ 

 50  &      0.82029  &      0.51093  &      0.47593  &      0.42089  &      0.49170     \\ 

 60  &      0.76513  &      0.47277  &      0.42939  &      0.37288  &      0.45195     \\ 

 70  &      0.69329  &      0.42913  &      0.37436  &      0.31560  &      0.40559     \\ 

 80  &      0.60022  &      0.38145  &      0.31155  &      0.25043  &      0.35401     \\ 

 82  &      0.57862  &      0.37174  &      0.29829  &      0.23678  &      0.34337     \\ 

 90  &      0.48131  &      0.33448  &      0.24491  &      0.18249  &      0.30185     \\ 

 92  &      0.45417  &      0.32607  &      0.23196  &      0.16958  &      0.29224     \\

\hline
\hline

\end{tabular}%

\end{table}

%% file: table_01orders_Breit_2p3_Z18-Z92.tex
\begin{table}[t]
\centering

\renewcommand{\arraystretch}{0.85}

\caption{\label{tab:IntEl01}
Nuclear-recoil effect on the $^2P_{3/2}$-state $g$ factor of B-like ions 
for the Coulomb and various effective potentials within the Breit approximation.
The contributions of zeroth [$A(\alpha Z)$, Eq.~(\ref{A_def})] and first [$B(\alpha Z)/Z$, Eq.~(\ref{B_def})] orders 
in the interelectronic interaction are presented together with their sum.
}

\begin{tabular}{l@{\quad\,}
                c@{\quad\,}
                S[table-format=-1.6,group-separator=]@{\quad\,}
                S[table-format=-1.6,group-separator=]@{\quad\,}
                S[table-format=-1.6,group-separator=]@{\quad\,}
                S[table-format=-1.6,group-separator=]@{\quad\,}
                S[table-format=-1.6,group-separator=]                 
               }
               
\hline
\hline

 $Z$   &  Term        
       &  {Coul}
       &  {CH  }
       &  {PZ  }
       &  {KS  }
       &  {LDF }  \\
        
\hline        

 18  &  $A$     &   -0.384348  &    -0.360555  &    -0.359640  &    -0.356128  &    -0.359513    \\ 
     &  $B/Z$   &    0.050930  &     0.033170  &     0.031857  &     0.027865  &     0.031836    \\ 
     &  $A+B/Z$ &   -0.333417  &    -0.327386  &    -0.327783  &    -0.328264  &    -0.327677    \\[1.5mm] 




 50  &  $A$     &   -0.349022  &    -0.342078  &    -0.341439  &    -0.340381  &    -0.341727    \\ 
     &  $B/Z$   &    0.016406  &     0.010219  &     0.009519  &     0.008418  &     0.009834    \\ 
     &  $A+B/Z$ &   -0.332616  &    -0.331859  &    -0.331920  &    -0.331963  &    -0.331893    \\[1.5mm] 






 92  &  $A$     &   -0.223800  &    -0.222179  &    -0.221182  &    -0.220520  &    -0.221820    \\ 
     &  $B/Z$   &    0.004937  &     0.003544  &     0.002521  &     0.001843  &     0.003177    \\ 
     &  $A+B/Z$ &   -0.218864  &    -0.218635  &    -0.218660  &    -0.218677  &    -0.218644    \\[1.5mm]

\hline
\hline

\end{tabular}%

\end{table}

%% file: table_B_01_total_2p3_Z18-Z92.tex
\begin{table}[t]
\centering

\renewcommand{\arraystretch}{1.1}

\caption{\label{tab:total}
Breit, QED, and total nuclear-recoil contributions to the $^2P_{3/2}$-state 
$g$ factor of B-like ions expressed in terms of the function $F(\alpha Z)$ which is defined by Eq.~(\ref{F_def}).
}

\begin{tabular}{l@{\qquad}
                S[table-format=-1.5(2),group-separator=]@{\qquad}
                S[table-format= 1.5(1),group-separator=]@{\qquad}
                S[table-format=-1.5(2),group-separator=]                 
               }
               
\hline
\hline

 $Z$   &  {$F_{\rm Breit}(\alpha Z)$}
       &  {$F_{\rm QED  }(\alpha Z)$}
       &  {$F_{\rm rec  }(\alpha Z)$}   \\
        
\hline        

 18  &     -0.3277(68)  &      0.00004(1)  &     -0.3276(68)     \\ 

 20  &     -0.3329(55)  &      0.00006(1)  &     -0.3328(55)     \\ 

 30  &     -0.3437(24)  &      0.00019(3)  &     -0.3435(24)     \\ 

 40  &     -0.3416(13)  &      0.00041(5)  &     -0.3411(13)     \\ 

 50  &    -0.33189(78)  &      0.00071(9)  &    -0.33118(78)     \\ 

 60  &    -0.31597(50)  &     0.00103(14)  &    -0.31493(52)     \\ 

 70  &    -0.29373(33)  &     0.00128(20)  &    -0.29244(38)     \\ 

 80  &    -0.26449(21)  &     0.00131(27)  &    -0.26318(35)     \\ 

 82  &    -0.25771(20)  &     0.00126(29)  &    -0.25645(35)     \\ 

 90  &    -0.22718(14)  &     0.00085(36)  &    -0.22633(39)     \\ 

 92  &    -0.21864(13)  &     0.00067(38)  &    -0.21797(40)     \\

\hline
\hline

\end{tabular}%

\end{table}

%% file: table_isotopes_2p3_Z18-Z92.tex
\begin{table}[t]
\centering

\renewcommand{\arraystretch}{1.1}

\caption{\label{tab:isotopes}
Nuclear recoil contribution to the $^2P_{3/2}$-state $g$ factor of selected B-like ions in the range $Z=18-92$.
}

\begin{tabular}{c@{\qquad}
                S[table-format=2.5]@{\qquad}
                S[table-format=-1.3(2)]              
               }
               
\hline
\hline
  
 Ion    &  {$(m/M)\cdot 10^6$}   
       &  {$\Delta g_{\rm rec}  \cdot 10^6$}   \\
        
\hline        

     $^{40}_{18}{\rm Ar}^{13+}$  &         13.7308  &      -4.499(93)     \\ 

     $^{40}_{20}{\rm Ca}^{15+}$  &         13.7311  &      -4.570(75)     \\ 

     $^{48}_{20}{\rm Ca}^{15+}$  &         11.4427  &      -3.809(63)     \\ 

    $^{120}_{50}{\rm Sn}^{45+}$  &         4.57628  &       -1.516(4)     \\ 

    $^{142}_{60}{\rm Nd}^{55+}$  &         3.86665  &       -1.218(2)     \\ 

    $^{208}_{82}{\rm Pb}^{77+}$  &         2.63826  &       -0.677(1)     \\ 

    $^{238}_{92}{\rm U }^{87+}$  &         2.30495  &       -0.502(1)     \\

\hline
\hline

\end{tabular}%

\end{table}